# Electrodynamics in material media

Bernhard Rothenstein
Department of Physics, "Politehnica" University of Timisoara, Romania

Aldo De Sabata
Faculty of Electronics and Telecommunications, "Politehnica" University of Timisoara, Romania

*Transformation equations for physical quantities that characterize plane electromagnetic wave propagation in transparent optical media are presented. The Doppler effect, and measurements performed by an observer moving with the wave are also considered. The results, not mentioned in introductory Physics textbooks, are simple to derive and could be used in undergraduate courses.*

## 1. Introduction

An electromagnetic wave propagates through empty space with velocity $c$, the upper limit of all physical velocities. It also propagates through non-conducting media with velocity

$$u = \frac{c}{n} = \frac{1}{\sqrt{\varepsilon\mu}} \qquad (1)$$

where $n$ stands for the *refractive index*, $\varepsilon$ for the *permittivity* and $\mu$ for the *permeability* of the medium. Because $n>1$ and $u<c$, we can define observers who move with velocities equal to or higher than $u$. We find a similar situation in the case of an acoustic wave where observers could move with supersonic velocities.[1]

In the non-conducting medium, we detect four fields: *the electric and the magnetic fields*, **E** and **B** and the *electric displacement* and *magnetic intensity fields* **D** and **H**, respectively. The physical quantities introduced in order to characterize the electric and the magnetic field ($\varepsilon,\mu$) and the fields defined above could have different values for different inertial observers in relative motion.

We could characterize the electromagnetic wave propagating through a medium by the *period* and the *frequency* of the electromagnetic oscillations taking place in the wave and by its *wavelength.*

The purpose of our paper is to present a simple and transparent approach to all relativistic aspects of electrodynamics in material media. The prerequisites are a knowledge of electricity and magnetism at an introductory level (Lorentz force)[2] and of the addition law of relativistic velocities[3].

## 2. Transformation equations for physical quantities introduced in order to characterize the electromagnetic field and the medium through which it propagates

The inertial reference frames involved in our approach are the rest frame of the medium $K(XOY)$ and a reference frame $K'(X'O'Y')$, both supposed to be inertial. Reference frame $K'(X'O'Y')$ moves with constant velocity $v$ in the positive direction of the overlapped $OX$ and $O'X'$ axes. The corresponding axes of the two frames are parallel to each other and, at the origin of time, the origins $O$ and $O'$ are instantly located at the same point in space.

We can derive the addition law of relativistic velocities without using the Lorentz-Einstein transformations for the space-time coordinates of the same event.[4,5,6] If a particle moves with velocity $u$ relative to $K(XOY)$ and with velocity $u'$ relative to the $K'(X'O'Y')$ reference frame then, in the positive direction of the overlapped $OX(O'X')$ axes, special relativity teaches us that equations

$$u = \frac{u' + v}{1 + \frac{u'v}{c^2}} \tag{2}$$

and

$$u' = \frac{u - v}{1 - \frac{vu}{c^2}} \tag{3}$$

relate them. It is worthwhile to emphasize that (3) can be obtained by solving (2) for $u'$ and vice-versa.

The considered *plane* and *polarized* electromagnetic wave propagates in the positive direction of the $OX(O'X')$ axes with the *phase velocities* $u$ and $u'$ relative to $K$ and $K'$, respectively. The vectors **E, B, D** and **H** are perpendicular to the direction of propagation in such a way that the vector products $\mathbf{E} \times \mathbf{B}$ and $\mathbf{D} \times \mathbf{H}$ are in the direction of propagation (Figure 1). In order to illustrate this fact we use the notations $\mathbf{E}_y, \mathbf{D}_y, \mathbf{B}_z$ and $\mathbf{H}_z$.

The law of electromagnetic induction[7] leads to the following relationship

$$E_y = uB_z \tag{4}$$

in $K(XOY)$ and

$$E_y' = u'B_z' \tag{5}$$

in $K'(X'O'Y')$.

The corresponding physical quantities transform as[8]

$$E_y' = \gamma(v)(E_y - vB_z) = \gamma(v)E_y\left(1 - \frac{v}{u}\right) = \frac{u'}{\gamma(v)(u'+v)}E_y = FE_y \tag{6}$$

$$B_y' = \gamma(v)\left(B_z - \frac{v}{c^2}E_y\right) = \gamma(v)B_z\left(1 - \frac{vu}{c^2}\right) = \frac{1}{\gamma(v)(1 + \frac{vu'}{c^2})} = GB_z. \tag{7}$$

The electric displacement field and the magnetic field intensity transform as

$$D_y' = \gamma(v)\left(D_y - \frac{v}{c^2}H_z\right) = \gamma(v)D_y\left(1 - \frac{vu}{c^2}\right) = \frac{1}{\gamma(v)\left(1 + \frac{u'v}{c^2}\right)}D_y = GD_y \tag{8}$$

$$H_z' = \gamma(v)(H_z - vD_y) = \gamma(v)(1 - \frac{v}{u})H_z = \frac{u'}{\gamma(v)(u'+v)}H_z = FH_z. \tag{9}$$

We emphasize the analogy between the equations introduced above (6)-(9) and the transformation equations for the space-time coordinates of the same event $x,t$ $(x',t')$[9]

$$\Delta x' = \gamma(v)(\Delta x - v\Delta t) = \gamma(v)\Delta x(1 - \frac{v}{u}) = F\Delta x \tag{10}$$

$$\Delta t' = \gamma(v)\left(\Delta t - \frac{v\Delta x}{c^2}\right) = G\Delta t \tag{10'}$$

and the transformation equations for the momentum and the energy $p, W$ $(p', W')$

$$p' = Fp \tag{11}$$

$$W' = GW. \tag{12}$$

We can say that momentum, electric field and magnetic field intensity are space like quantities, whereas energy, electric field displacement and magnetic field are time like ones.

**3. Relativistic transformations for the optical constants of media**

The optical properties of a transparent medium are $\varepsilon$, $\mu$ and the refraction index $n = \dfrac{c}{u}$ in $K(XOY)$ and $\varepsilon'$, $\mu'$ and $n' = \dfrac{c}{u'}$ in $K'(X'O'Y')$.

Combining (8) and (6) we obtain

$$\varepsilon' = \frac{D_y'}{E_y'} = \frac{G}{F}\varepsilon \qquad (13)$$

the transformation equation for the permittivity. Combining (7) and (9) we obtain

$$\mu' = \frac{B_z'}{H_z'} = \frac{G}{F}\mu \qquad (14)$$

the transformation equation for the permeability. As we see

$$\frac{\varepsilon'}{\mu'} = \frac{\varepsilon}{\mu} \qquad (15)$$

is a relativistic invariant ($Z = \sqrt{\dfrac{\mu}{\varepsilon}}$ is the wave impedance of the medium).

**4. Doppler Effect in the electromagnetic wave propagating in a medium**

When we speak about a physical quantity, it is advisable to mention the observer who performs the measurement, when and where he performs the measurement and the measuring devices he uses.

Consider that at the origin of time ($t=t'=0$) observers $R_0(0,0)$ and $R_0'(0,0)$, the first located at the origin $O$ of $K(XOY)$ and at rest relative to it, the second located at the origin $O'$ of $K'(X'O'Y')$ and at rest relative to it, are instantly located at the same point in space. At that very moment, the front of the wave arrives at this point. During a period $T$ of the electromagnetic oscillations taking place in the wave, the front of the wave travels a distance

$$\lambda = uT \qquad (16)$$

in $K(XOY)$ and

$$\lambda' = u'T' \qquad (17)$$

in $K'(X'O'Y')$. By definition $\lambda$ and $\lambda'$ represent the wavelength of the wave in $K(XOY)$ and in $K'(X'O'Y')$ respectively. Under such conditions, the front of the wave generates the events $M(\lambda, 0, \dfrac{\lambda}{u}) = M(uT, 0, T)$ in $K(XOY)$ and $M'(\lambda', 0, \dfrac{\lambda'}{u'}) = M'(u'T', 0, T')$ in $K'(X'O'Y')$. The Lorentz-Einstein transformations for the space-time coordinates lead to

$$\lambda' = \gamma(v)\lambda(1 - \frac{v}{u}) = F\lambda \qquad (18)$$

$$T' = \gamma(v)T(1 - \frac{vu}{c^2}) = GT. \qquad (19)$$

In the experiment described above, the involved observers measure the period of the electromagnetic oscillations taking place in the wave and the propagation velocity of the wave and then determine the wavelength. The oscillations taking place in the wave serve as clocks.

The division of equations (18), (19) and the use of (16), (17) prove that the phase velocities add as the velocities of subluminal particles do.

The transformation factors $F$ and $G$ become in the case of an electromagnetic wave propagating in empty space ($u=u'=c$)

$$F_c = G_c = \sqrt{\frac{1-\frac{v}{c}}{1+\frac{v}{c}}} \ . \tag{20}$$

## 5. Energy density and Poynting vector in the electromagnetic wave propagating in a medium

We define the energy density ($w,w'$) as energy stored in a unit volume. Electromagnetic theory leads to[10]

$$w = \varepsilon E^2 = \frac{B^2}{\mu} \tag{21}$$

in $K(XOY)$ and to

$$w' = \varepsilon E'^2 = \frac{B'^2}{\mu'} \tag{22}$$

resulting that it transforms as

$$w' = \frac{\varepsilon'}{\varepsilon}\frac{E'^2}{E^2}w = FGw. \tag{22'}$$

The Poynting vector ($\Pi$, $\Pi'$) defined as energy carried through the unit surface normal to the direction of propagation and during the unit time interval has the magnitude

$$\Pi = wu = EH = \sqrt{\frac{\varepsilon}{\mu}}E^2 \tag{23}$$

in $K(XOY)$ and

$$\Pi = w'u' = E'H' = \sqrt{\frac{\varepsilon'}{\mu'}}E'^2 \tag{24}$$

in $K'(X'O'Y')$. It transforms as

$$\Pi' = \frac{E'^2}{E^2}\Pi = F^2\Pi \ . \tag{25}$$

## 6. The electromagnetic wave propagating through a medium as a velocity filter

Consider a particle with positive electric charge $q$ traveling with velocity $w'$ in the positive direction of $OX(O'X')$ axes. An electric force acts upon it

$$F_{e,y}' = qE_y' \tag{26}$$

and a magnetic force

$$F_{m,y}' = qw'B_z' \tag{27}$$

the first showing in the positive direction of the $O'Y'$ axis, the second in its negative direction. The resultant force is

$$F_y' = q(E_y' - w'B_z') \ . \tag{28}$$

As we see, if the particle moves with velocity $w'=u'$ it is not deviated from its initial direction of motion. Charged particles moving with subluminal velocities ($w'<u'$) are deviated in the positive direction of the $O'Y'$ axis whereas those moving with superluminal velocities ($w'>u'$) will be deviated in its negative direction.

## 7. The observer who moves with the wave

McTavish[11] considers a special observer $R_{u'_{x=0}}$ 'who moves with velocity $w=u$ relative to $K(XOY)$ and is at rest relative to $K'(X'O'Y')$ ($u'=0$). For such an observer

$$F_{u,u'=0} = 0 \tag{29}$$

and

$$G_{u,u'=0} = \sqrt{1-\frac{u^2}{c^2}} \ . \tag{30}$$

The result is that, from the point of view of that observer, all the physical quantities, which transform via the factor F are zero. He measures a zero electric field ($E'=0$), a zero magnetic field intensity ($H'=0$)) but a different from zero magnetic field

$$B' = \sqrt{1-\frac{u^2}{c^2}}B \tag{31}$$

and a different from zero electric displacement field

$$D' = \sqrt{1-\frac{u^2}{c^2}}D \ . \tag{32}$$

He also detects a zero energy density ($w'=0$) and a zero Poynting vector.

## 7. Conclusions

We present transformation equations for the physical quantities, which characterize the optical properties of a transparent medium as well as for the results of the measurements performed by observers in relative motion and located in the medium through which an electromagnetic wave propagates. The results are simple to derive and could be the subject of a workshop in an undergraduate Physics course.
Results not mentioned in introductory Physics textbooks are presented and the relativistic invariance of $\frac{\varepsilon}{\mu}$ is proved. The medium is supposed to be linear, isotropic and homogeneous, responding instantaneously to the action of the propagating electromagnetic wave.